# A STAGED MUON ACCELERATOR FACILITY FOR NEUTRINO AND COLLIDER PHYSICS*

Jean-Pierre Delahaye, SLAC, Menlo Park, California
Charles Ankenbrandt, Stephen Brice, Alan David Bross, Dmitri Denisov, Estia Eichten, Stephen Holmes, Ronald Lipton, David Neuffer, Mark Alan Palmer, Fermilab, Batavia, Illinois
S. Alex Bogacz, JLAB, Newport News, Virginia
Patrick Huber, Virginia Polytechnic Institute and State University, Blacksburg
Daniel M. Kaplan, Pavel Snopok, Illinois Institute of Technology, Chicago, Illinois
Harold G. Kirk, Robert B. Palmer, BNL, Upton, Long Island, New York
Robert D. Ryne, LBNL, Berkeley, California

## Abstract

Muon-based facilities offer unique potential to provide capabilities at both the Intensity Frontier with Neutrino Factories and the Energy Frontier with Muon Colliders. They rely on a novel technology with challenging parameters, for which the feasibility is currently being evaluated by the Muon Accelerator Program (MAP). A realistic scenario for a complementary series of staged facilities with increasing complexity and significant physics potential at each stage has been developed. It takes advantage of and leverages the capabilities already planned for Fermilab, especially the strategy for long-term improvement of the accelerator complex being initiated with the Proton Improvement Plan (PIP-II) and the Long Baseline Neutrino Facility (LBNF). Each stage is designed to provide an R&D platform to validate the technologies required for subsequent stages. The rationale and sequence of the staging process and the critical issues to be addressed at each stage, are presented.

## INTRODUCTION

The major discoveries of the large flavour mixing angle $\theta_{13}$ at Daya Bay in China and of the Higgs boson by LHC at CERN dramatically modified the Particle Physics landscape. Although the Higgs discovery corresponds to a splendid confirmation of the Standard Model (SM) and no sign of physics Beyond Standard Model (BSM) has (yet) been detected at LHC, BSM physics is necessary to address basic questions which the SM cannot, especially dark matter, dark energy, matter-antimatter asymmetry and neutrino mass. Therefore the quest for BSM physics is a high priority for the future of High Energy Physics. It requires facilities at both the high energy and high intensity frontiers. Neutrino oscillations are irrefutable evidence for BSM physics with the potential to probe up to extremely high energies. Neutrino Factories with an intense and well defined flux of neutrinos from muon decay provide an ideal tool for high precision flavour physics at the intensity frontier. At the energy frontier, a multi-TeV lepton collider will be necessary as a precision facility to complement the LHC, for physics beyond the Standard Model if and when such physics is confirmed.

___________________
* Work supported by the U.S. Dept. of Energy under contracts DE-AC02-07CH11359 and DE-AC02-76SF00515.

## THE BEAUTY AND CHALLENGES OF MUON-BASED FACILITIES

Muon-based facilities [1] offer the unique potential to provide the next generation of capabilities and world-leading experimental support spanning physics at both the Intensity and Energy Frontiers. Building on the foundation of PIP-II and its successor stages at FNAL [2], muon accelerators can provide the next step with a high-intensity and precise source of neutrinos to support a world-leading research program in neutrino physics. Furthermore, the infrastructure developed to support such an Intensity Frontier research program can also enable the return of the U.S. high energy physics program to the Energy Frontier. This capability would be provided in a subsequent stage of the facility that would support one or more Muon Colliders, which could operate at center-of-mass energies from the Higgs resonance at 126 GeV up to the multi-TeV scale.

An ensemble of facilities built in stages is made possible by the strong synergies between Neutrino Factories and Muon Colliders, both of which require a high power proton source and target for muon generation followed by similar front-end and ionization cooling channels. These muon facilities rely on a number of systems with conventional technologies whose required operating parameters exceed the present state of the art as well as novel technologies unique to muon colliders. An R&D program to evaluate the feasibility of these technologies is being actively pursued within the framework of the U.S. Muon Accelerator Program (MAP) [3] with impressive R&D results already achieved and a definitive answer expected by 2020 as to whether or not muon-based facilities built in stages can be realistically contemplated.

## RATIONALE FOR A STAGED APPROACH

The feasibility of the technologies required for Neutrino Factories and/or Muon Colliders must be validated before a facility based upon these could be proposed. Such validation is usually made in dedicated test facilities which are rather expensive to build and to operate over several years. They are therefore difficult to justify and fund, given especially that they are usually useful only for technology development rather than for physics.

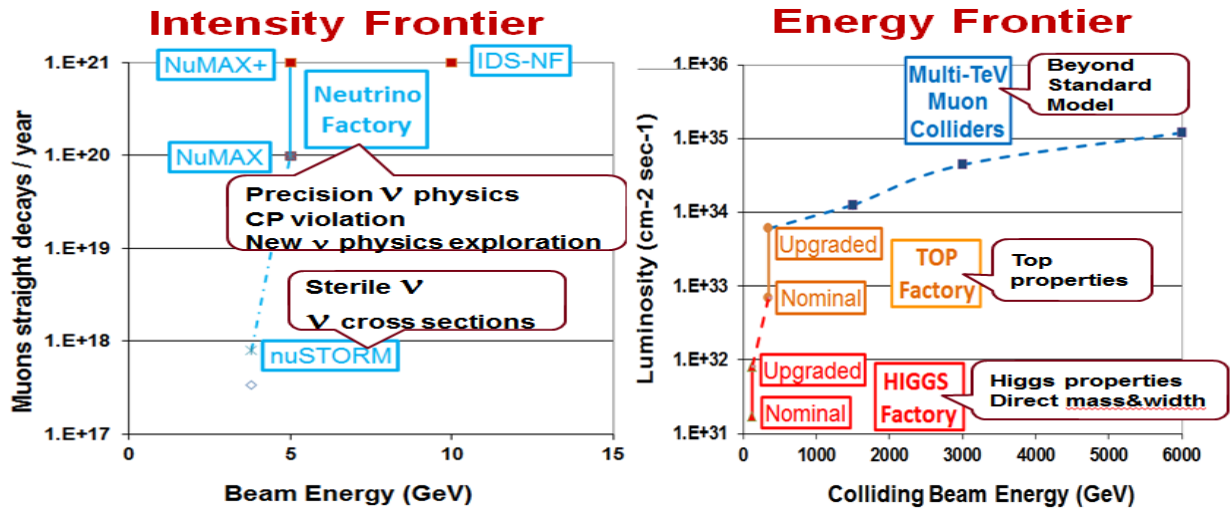

Figure 1: Performance and physics of muon-based facilities over two frontiers and a wide energy range.

An alternative approach is proposed here. It consists of a series of facilities built in stages, where each stage offers:

- Unique physics capabilities such that the facility obtains support and is funded.
- In parallel with the physics program, integration of an R&D platform to develop, test with beam, validate and get operational experience with a new technology that is necessary for the following stages.
- Construction of each stage as an add-on to the previous stages, extensively reusing the equipment and systems already installed, such that the additional budget of each stage remains affordable.

Such a staging plan [4,5] thus provides clear decision points before embarking upon each subsequent stage. It is especially attractive at FNAL building on, and taking advantage of, existing or proposed facilities, specifically:

- Existing tunnels and other conventional facilities;
- An advanced proton complex based on PIP-II as the MW-class proton driver for muon generation;
- SURF as developed for the LBNF detector, which could then house the detector for a long-baseline Neutrino Factory.

The staging plan consists of a series of facilities with increasing complexity, each with performance characteristics providing unique physics reach (Fig. 1) and from which parameters are summarized in Table 1:

- **nuSTORM** [6]: a short-baseline Neutrino Factory-like facility enabling a definitive search for sterile neutrinos, as well as neutrino cross-section measurements that will ultimately be required for precision measurements at any long-baseline experiment.
- **NuMAX** (**N**eutrinos from **M**uon **A**ccelerator Comple**X**): a long-baseline 5 GeV Neutrino Factory, optimized for a detector at SURF 1300 km from Fermilab, providing a precise and well-characterized neutrino source that exceeds the capabilities of conventional superbeams.
- **NuMAX**+: a full-intensity Neutrino Factory, upgraded from NuMAX, with performances similar to IDS-NF [7] as the ultimate source to enable precision CP-violation measurements and potential exploration of new physics in the neutrino sector.
- **Higgs Factory**: a collider capable of providing between 3500 (startup) and 13,500 Higgs events per year ($10^7$ sec) with exquisite energy resolution enabling direct Higgs mass and width measurements.
- **Multi-TeV Collider**: if warranted by LHC results, a multi-TeV Muon Collider, with an ultimate energy reach of 6 to 10 TeV, likely offers the best performance and least cost and power consumption for any lepton collider operating in the multi-TeV regime (Fig. 2).

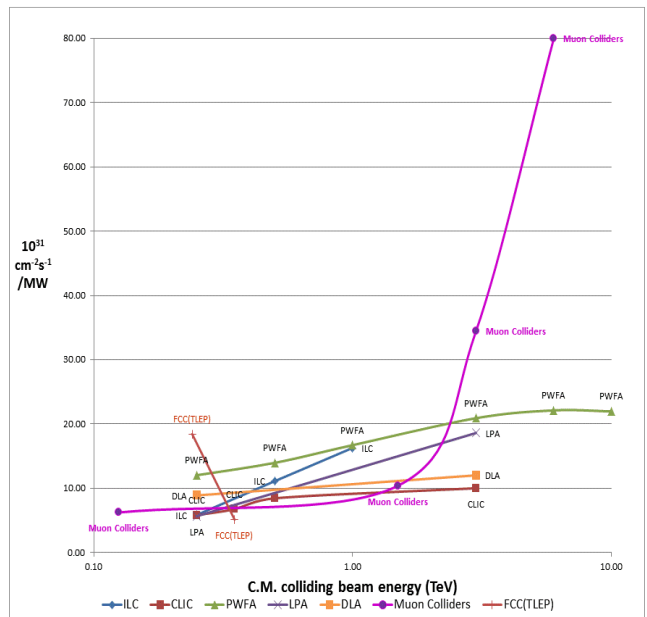

Figure 2: Luminosity per wall plug power consumption as figure of merit of the various lepton collider technologies [5].

Table 1: Main Parameters of the Various Muon-Based Facilities in a Phased Approach

| System | Parameters | Unit | nuSTORM | NuMAX Commissioning | NuMAX | NuMAX+ |
|---|---|---|---|---|---|---|
| Performance | $\nu_e$ or $\nu_\mu$ to detectors/year | - | $3\times10^{17}$ | $4.9\times10^{19}$ | $1.8\times10^{20}$ | $5.0\times10^{20}$ |
| | Stored $\mu^+$ or $\mu^-$/year | - | $8\times10^{17}$ | $1.25\times10^{20}$ | $4.65\times10^{20}$ | $1.3\times10^{21}$ |
| Detector | Far Detector: Type | | SuperBIND | MIND / Mag LAr | MIND / Mag LAr | MIND / Mag LAr |
| | Distance from Ring | km | 1.9 | 1300 | 1300 | 1300 |
| | Mass | kT | 1.3 | 100 / 30 | 100 / 30 | 100 / 30 |
| | Magnetic Field | T | 2 | 0.5-2 | 0.5-2 | 0.5-2 |
| | Near Detector: Type | | SuperBIND | Suite | Suite | Suite |
| | Distance from Ring | m | 50 | 100 | 100 | 100 |
| | Mass | kT | 0.1 | 1 | 1 | 2.7 |
| | Magnetic Field | T | Yes | Yes | Yes | Yes |
| Neutrino Ring | Ring Momentum | GeV/c | 3.8 | 5 | 5 | 5 |
| | Circumference (C) | m | 480 | 737 | 737 | 737 |
| | Straight section | m | 184 | 281 | 281 | 281 |
| | Number of bunches | - | | 60 | 60 | 60 |
| | Charge per bunch | | $1\times10^9$ | 6.9 | 26 | 35 |
| Acceleration | Initial Momentum | GeV/c | - | 0.25 | 0.25 | 0.25 |
| | Single-pass Linacs | GeV/c | - | 1.0, 3.75 | 1.0, 3.75 | 1.0, 3.75 |
| | | MHz | - | 325, 650 | 325, 650 | 325, 650 |
| | Repetition | Hz | - | 30 | 30 | 60 |
| Cooling | | | No | No | Initial | Initial |
| Proton Driver | Proton Beam Power | MW | 0.2 | 1 | 1 | 2.75 |
| | Proton Beam | GeV | 120 | 6.75 | 6.75 | 6.75 |
| | Protons/year | $1\times10^{21}$ | 0.1 | 9.2 | 9.2 | 25.4 |
| | Repetition | Hz | 0.75 | 15 | 15 | 15 |

| Muon Collider Parameters | | Higgs Factory | | Top Threshold Options | | Multi-TeV Baselines | | |
|---|---|---|---|---|---|---|---|---|
| Parameter | Units | Startup Operation | Production Operation | High Resolution | High Luminosity | | | Accounts for Site Radiation Mitigation |
| CoM Energy | TeV | 0.126 | 0.126 | 0.35 | 0.35 | 1.5 | 3.0 | 6.0 |
| Avg. Luminosity | $10^{34}$cm$^{-2}$s$^{-1}$ | 0.0017 | 0.008 | 0.07 | 0.6 | 1.25 | 4.4 | 12 |
| Beam Energy Spread | % | 0.003 | 0.004 | 0.01 | 0.1 | 0.1 | 0.1 | 0.1 |
| Higgs* or Top† Production/10$^7$sec | | 3,500* | 13,500* | 7,000† | 60,000† | 37,500* | 200,000* | 820,000* |
| Circumference | km | 0.3 | 0.3 | 0.7 | 0.7 | 2.5 | 4.5 | 6 |
| No. of IPs | | 1 | 1 | 1 | 1 | 2 | 2 | 2 |
| Repetition Rate | Hz | 30 | 15 | 15 | 15 | 15 | 12 | 6 |
| $\beta^*$ | cm | 3.3 | 1.7 | 1.5 | 0.5 | 1 (0.5-2) | 0.5 (0.3-3) | 2.5 |
| No. muons/bunch | $10^{12}$ | 2 | 4 | 4 | 3 | 2 | 2 | 2 |
| No. bunches/beam | | 1 | 1 | 1 | 1 | 1 | 1 | 1 |
| Norm. Trans. Emittance, $\varepsilon_{TN}$ | $\pi$ mm-rad | 0.4 | 0.2 | 0.2 | 0.05 | 0.025 | 0.025 | 0.025 |
| Norm. Long. Emittance, $\varepsilon_{LN}$ | $\pi$ mm-rad | 1 | 1.5 | 1.5 | 10 | 70 | 70 | 70 |
| Bunch Length, $\sigma_s$ | cm | 5.6 | 6.3 | 0.9 | 0.5 | 1 | 0.5 | 2 |
| Proton Driver Power | MW | 4‡ | 4 | 4 | 4 | 4 | 4 | 1.6 |

## A POSSIBLE STAGED SCENARIO

The staged scenario is based on a progressive implementation of facilities with increasing complexity by adding systems to the previously installed systems. It takes advantage of the strong synergies between Neutrino Factory and Muon Collider. Both require:
- A proton driver producing a high-power multi-GeV bunched proton beam.
- A pion production target operating in a high-field solenoid. A solenoid confines the pions radially, guiding them into a decay channel.
- A "front end" consisting of a solenoid focused $\pi\rightarrow\mu$ decay channel, followed by a system of RF cavities to capture the muons longitudinally and phase rotate them into a suitable bunch train.
- A cooling channel that uses ionization cooling to reduce the phase space occupied by the beam from the initial volume at the exit of the front end to what is required by the facility.
- A series of fast acceleration stages to take the muon beams to the relevant factory or collider energies.

### Neutrino Factory: NuMAX

The block diagram of a Neutrino Factory is displayed in Fig. 3a. It is based on an extension of the PIP-II linac accelerating the proton beam in two stages up to 3 GeV and further accelerated by a straight dual 650 MHz linac before hitting the target for pion production. The muons produced by pion decay, captured and bunched in the front end are recirculated to the dual linac for further acceleration up to 5 GeV as required by NuMAX. The dual use linac concept accelerating both the proton and muon beams provides an opportunity for considerable savings. It requires initial cooling to match muon beam emittances to the linac acceptances at the 325 and 650 MHz RF standards adopted by the FNAL PIP-II program. The initial cooling specifications result from a cost optimization as the best trade-off between linac and cooling. The Neutrino Factory complex would be built in

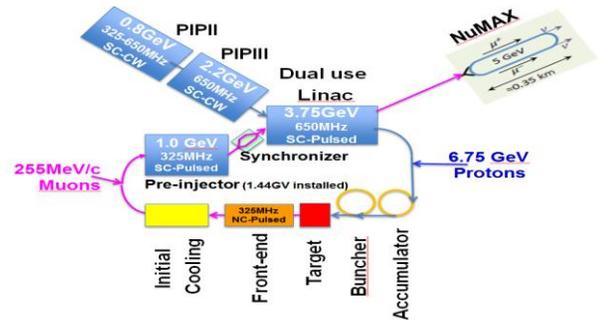

a) Layout of a Muon based Neutrino factory

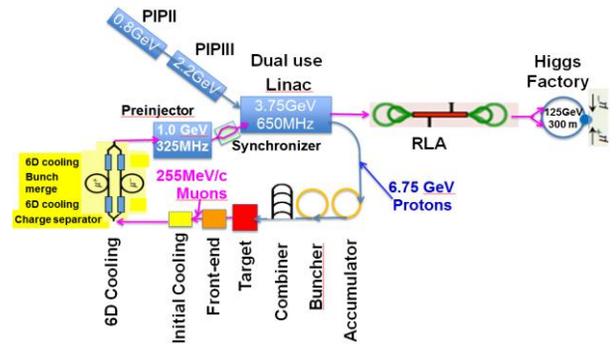

b) Layout of a Muon based Higgs factory

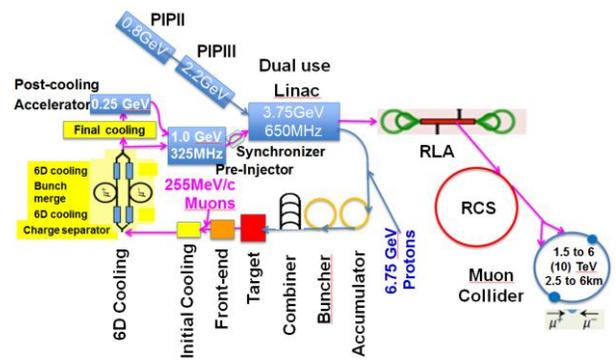

c) Layout of a multi-TeV Muon Collider

Figure 3: Evolution of Muon based complex from Neutrino Factory to Muon Collider.

phases with physics performance summarized in Fig. 4:

A **commissioning phase** based on a limited proton beam power of 1 MW on the muon production target and without any cooling for an early and realistic start with technology as conventional as possible, while already providing very attractive physics reach.

**NuMAX** upgraded from the commissioning phase by adding a limited amount of 6D cooling of both species of muon beams by a factor 5 in each transverse plane and a factor 2 in the longitudinal plane. This so-called initial cooling matches muon beam emittances at the exit of the front-end to the acceptances of the 3.75 GeV 650 MHZ dual linac and improves the performance by a factor 3.5.

**NuMAX+**: upgraded progressively from NuMAX by multiplying the proton beam power on target when it becomes available, and upgrading correspondingly the detector for a performance similar to the IDS-NF [7].

At each phase of the facility used as an R&D platform in parallel with physics, the muon beam at the front end exit is ideal to develop, test and validate the cooling required by the next phase, namely the initial cooling in nuSTORM and/or the NuMAX commissioning phase with $10^8$ to $10^9$ muon bunches and the 6D cooling at NuMAX+ with a $10^{12}$ muon beam.

### Higgs and/or Top Factory

All systems installed for NuMAX+ would be re-used for an upgrade of the facility to a low energy Muon Collider such as a Higgs or Top Factory (Fig. 3b). A 6D ionisation cooling reducing further the muon beam emittances by a factor 10 in both transverse planes and 15 in the longitudinal plane after validation during the previous NuMAX+ phase would be implemented as well as an RLA with multiple passes for efficient beam acceleration.

### Multi-TeV Muon Collider

Again all systems installed for a low energy collider would be re-used for an upgrade to a multi-TeV Muon Collider (Fig. 3c). A final cooling reducing the transverse emittances by another factor 15 as required for high luminosity of the collider will be added as well as a Rapid Cycling Synchrotron for further fast beam acceleration.
A complex integrating all of the above facilities in a staged approach on the FNAL site is shown in Figure 5.

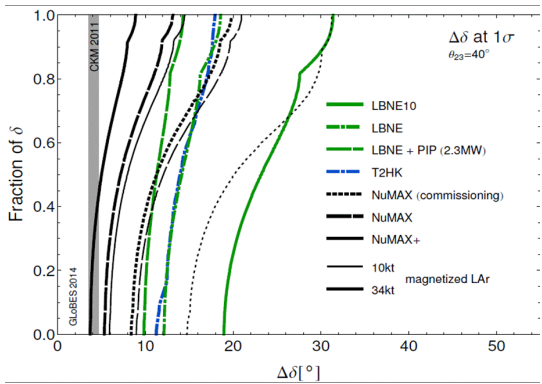

Figure 4: NuMAX CP violating physics performance.

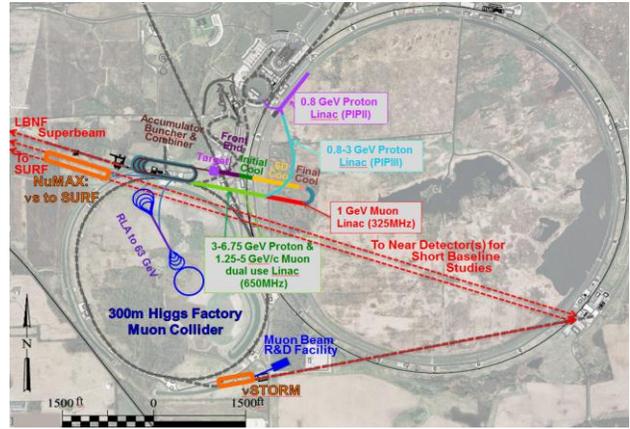

Figure 5: Footprint of Neutrino Factory and Higgs Factory Muon Collider facilities on the FNAL site. A 6 TeV Muon Collider would fit in a tunnel with the size of the Tevatron but deep underground.

## CRITICAL ISSUES AND R&D TO ADDRESS THEIR FEASIBILITY

The muon-based facilities rely on a number of systems with conventional technologies whose required operating parameters exceed the present state of the art as well as novel technologies unique to muon colliders. An R&D program to evaluate the feasibility of these technologies is being actively pursued within the framework of the U.S. Muon Accelerator Program (MAP) [3]. The critical issues of each stage include:

### For a Neutrino Factory (NuMAX)

- **A high-power proton linac and target station** of initially 1 MW upgradable to a few MW: The initial beam power significantly mitigates that issue by allowing use of already existing technologies. The maximum power, currently 4 MW, is being evaluated and possibly reduced to 2.3MW as presently envisaged for LBNF at FNAL thus maximizing synergies.
- **A 15–20 T capture solenoid**
- **RF accelerating gradient in low frequency (325–975 MHz) structures immersed in high magnetic field** as required for the front end and ionization cooling sections; promising results pointing towards solutions have recently been obtained in the MuCool Test Area (MTA) at Fermilab with accelerating gradients of 20 MV/m achieved without significant deterioration in a 0–5 T magnetic field range.

### For a Sub-TeV Collider

- **6D ionization cooling:** A Muon Cooling Ionization Experiment (MICE) [8] is presently being built at RAL to demonstrate the feasibility of ionization cooling and validate the simulation codes.
- **Recirculating Linear Accelerators (RLA)** with multiple passes: A dogbone RLA with 2-pass arcs is proposed to be built and operated at Jefferson Lab as a specific proof-of-concept electron test facility, JEMMRLA (JLab Electron Model of Muon RLA).

- **Collider ring design and machine–detector interface (MDI)** including absorbers for the decay products of the muon beams
- **Detector operation in a severe background environment** caused by muon decays around the ring: Much of the background is composed of soft particles with distinctive travel durations. Nanosecond time resolution has been shown to reduce the background by three orders of magnitude.

*For a Multi-TeV Collider*

- In the final cooling system, **very high-field (> 30 T) solenoids** utilizing high temperature superconducting (HTS) coils: A world record of 15 T on-axis field with YBCO superconductor has recently been achieved as well as a breakthrough in HTS cable performance with BSCCO-2212 conductor. A hybrid magnet prototype with HTS insert and anticipated performance in the required range is envisaged.
- **NC or SC magnets with ramp rates in the 1 to 10 T/ms range** in the Rapid Cycling Synchrotron (RCS): Concepts and prototypes of magnets and power supply systems are being developed.
- **Collider, MDI and Detector with high energy backgrounds** due to the high energy operation

## TENTATIVE SCHEDULE

Three major items on the critical path have been clearly identified: i) MAP technology feasibility study with results expected by 2020 if resources are made available, ii) 6D cooling with beam validation by 2025 possibly using nuSTORM as an R&D platform in complement to the cooling feasibility demonstration at MICE [8], iii) Proton beam power availability from PIP-II at 0.8 GeV by 2024 and somewhat later at 3 GeV. A technically limited schedule with informed decision by validation at each stage of the technology required by the next stage could then be defined, tentatively by the beginning of the next decade for a Neutrino Factory and by the middle of that decade for a Muon Collider (Fig. 6).

nuSTORM as the first stage could be launched as soon as funding is available since it does not require any technology development. A first beam could be provided by the end of the decade. In parallel with physics, it would be extremely useful for development of muon based technology and would provide an R&D platform to test and validate the novel ionization cooling.

NuMAX commissioning phase which does not require any cooling could be launched when the MAP technology feasibility has been demonstrated. It would initially be driven by a proton beam power of 1 MW. It would be upgraded progressively by adding cooling when validated and by increasing the beam power on target as available.

A multi-TeV Muon Collider could then be launched if and when required by physics for studies beyond the Standard Model by using the injector system previously built and adding the necessary final cooling, acceleration system and collider ring.

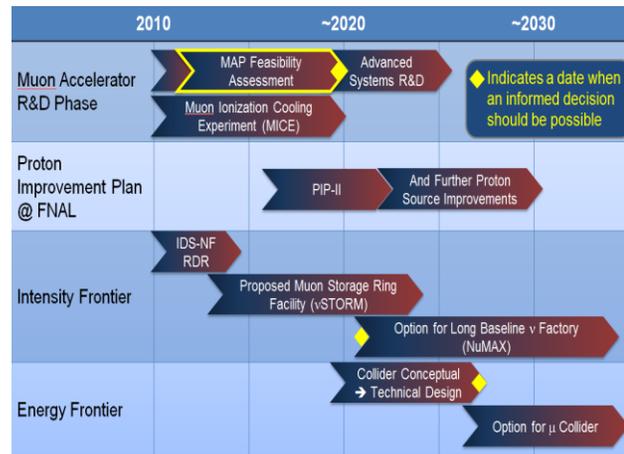

Figure 6: Tentative muon based facilities timeline with informed decisions dates.

## CONCLUSION

A staging approach for muon-based facilities with physics interest and technology validation at each phase has been developed as an attempt to find a realistic scenario taking advantage of the present and proposed facilities at Fermilab and to drive the R&D to demonstrate its feasibility. Each stage is built as an addition to the previous stages, reusing as much as possible the systems already installed such that the additional budget of each stage remains affordable. Thanks to the great synergies between Neutrino Factory and Muon Collider, these facilities are complementary and allow capabilities and world-leading experimental support spanning physics at both the Intensity and Energy Frontiers. Rather than building an expensive test facility without physics use, the approach uses each stage as an R&D platform at which to test and validate the technology required by the following stage. The critical issues of this novel and promising technology are thus addressed in the most efficient and practical way within reasonable funding and scheduling constraints. As deduced from the components on the critical path and a technically limited schedule, such a staging approach with integrated R&D allows informed decisions by early next decade about Neutrino Factories at the Intensity Frontier and by the middle of the next decade about Muon Colliders at the Energy Frontier.